\documentclass[conference,letterpaper]{preamble/IEEEtran}

\usepackage[utf8]{inputenc} 
\usepackage[T1]{fontenc}
\usepackage{url}
\usepackage{ifthen}
\usepackage{cite}
\usepackage[cmex10]{amsmath} 
\usepackage{graphicx,amssymb,amsfonts,adjustbox,arydshln,mathtools,balance,cite}
\usepackage{eqnarray}
\usepackage{mathabx,algorithmic,setspace}
\usepackage[ruled]{algorithm2e}
\usepackage{booktabs, makecell, multirow, tabularx}

\usepackage{tikz,pgf,pgffor,pgfplots,pgfplotstable,subfigure}
\usetikzlibrary{shapes,arrows.meta,positioning,shapes.geometric,fit,backgrounds}
\pgfplotsset{compat=1.17} 

\usepackage{xpatch}
\makeatletter
\xpatchcmd{\@thm}{\thm@headpunct{.}}{\thm@headpunct{}}{}{}
\makeatother

\makeatletter
\def\bstctlcite{\@ifnextchar[{\@bstctlcite}{\@bstctlcite[@auxout]}}
\def\@bstctlcite[#1]#2{\@bsphack
  \@for\@citeb:=#2\do{%
    \edef\@citeb{\expandafter\@firstofone\@citeb}%
    \if@filesw\immediate\write\csname #1\endcsname{\string\citation{\@citeb}}\fi}%
  \@esphack}
\makeatother

\mathchardef\mhyphen="2D

\newcommand{\C}{\mathcal C}

\newcommand{\B}{\mathcal B}

\newcommand{\Rn}{\mathbb{R}^n}
\newcommand{\F}{\mathbb F}

\newcommand{\Zn}{\mathbb{Z}^n}

\renewcommand{\H}{\mathbf H}

\newcommand{\G}{\mathbf G}
\newcommand{\D}{\mathbf D}
\newcommand{\T}{\mathbf T}

\newcommand{\mA}{\mathbf A}
\newcommand{\mB}{\mathbf B}
\newcommand{\mC}{\mathbf C}

\newcommand{\mX}{\mathbf X}

\newcommand{\bb}{\mathbf b}

\newcommand{\bu}{\mathbf u}

\newcommand{\h}{\mathbf h}
\newcommand{\w}{\mathbf w}
\newcommand{\x}{\mathbf x}
\newcommand{\y}{\mathbf y}
\newcommand{\z}{\mathbf z}

\newcommand{\zero}{\mathbf 0}

\newcommand{\Gs}{\mathbf{G}_{\mathrm{s}}}
\newcommand{\Ls}{\Lambda_{\mathrm{s}}}
\newcommand{\Lc}{\Lambda_{\mathrm{c}}}

\newcommand{\thx}{\widehat{\widetilde{\mathbf x}}}
\newcommand{\tx}{\widetilde{\mathbf x}}
\newcommand{\hx}{\widehat{\mathbf x}}
\newcommand{\tH}{\widetilde{\mathbf H}}
\newcommand{\hu}{\widehat{\mathbf u}}

\newcommand{\starmod}{\textrm{ mod}^* }

\newcommand{\modtwo}{\textrm{mod}_2 }
\renewcommand{\mod}{\textrm{ mod }}

\newcommand{\Dec}{\textrm{Dec}}

\newcommand{\SNR}{\textrm{SNR}}

\newcommand{\VNR}{\textrm{VNR}}

\newcommand{\dB}{\textrm{ dB}}

\newcommand{\eb}{E_{\mathrm b}}
\newcommand{\no}{N_0}

\newcommand{\ebnoshort}{\eb/\no}
\newcommand{\var}{\sigma^2}

\newtheorem{definition}{Definition}

\newtheorem{proposition}{Proposition}

\renewcommand{\IEEEQED}{\IEEEQEDopen}

\usepackage{hyperref}

\IEEEoverridecommandlockouts

\IEEEaftertitletext{\vspace{-0.4\baselineskip}}
\begin{document}
\title{Encoding and Decoding {C}onstruction D' Lattices \\ for Power-Constrained Communications} 

\author{%
  \IEEEauthorblockN{Fan Zhou\IEEEauthorrefmark{1}, Arini Fitri\IEEEauthorrefmark{7}, Khoirul Anwar\IEEEauthorrefmark{7}, and Brian M.~Kurkoski\IEEEauthorrefmark{1}}
  \IEEEauthorblockA{\IEEEauthorrefmark{1}%
                    School of Information Science, Japan Advanced Institute of Science and Technology}
 \IEEEauthorblockA{\IEEEauthorrefmark{7}%
                The University Center of Excellence for AICOMS, Telkom University}
\thanks{This work was supported by JSPS Kakenhi Grant Number JP 19H02137. This work is also the output of the ASEAN IVO project, PATRIOT-41R-Net,  financially supported by NICT, Japan.
}

}
\maketitle
\begin{abstract}
This paper focuses on the encoding and decoding of {C}onstruction D' coding lattices that can be used with shaping lattices for power-constrained channels.
Two encoding methods and a decoding algorithm for {C}onstruction D' lattices are given. 
A design of quasi-cyclic low-density parity-check (QC-LDPC) codes to form {C}onstruction D' lattices is presented. This allows construction of nested lattice codes which are good for coding, good for shaping, and have low complexity encoding and decoding. Numerical results of using $E_8$, $BW_{16}$ and Leech lattices for shaping a {C}onstruction D' lattice indicate that the shaping gains $0.65\dB$, $0.86\dB$ and $1.03\dB$ are preserved, respectively. 

\end{abstract}


\section{Introduction} \label{section:introduction}
Lattices are a natural fit for wireless communications because they provide reliable transmission using real-valued algebra and higher transmit power efficiency than conventional constellations. Lattices also form an important component of compute-and-forward relaying~\cite{Nazer-it11}, which provides high throughput and high spectral efficiency. For power-constrained communications, nested lattice codes constructed by a coding lattice $\Lc$ and a shaping lattice $\Ls$ can achieve the additive white {G}aussian noise (AWGN) channel capacity~\cite{Erez-it04}, if $\Lc$ is \emph{channel-good} and the {V}oronoi region of $\Ls$ is hyperspherical, using dithering and minimum mean-square error (MMSE) scaling techniques. The shaping gain measures the power reduction, with a $1.53\dB$ theoretic limit. The well-known $E_8$ lattice was employed for shaping {C}onstruction D lattices~\cite{buglia-arxiv20}. Lattices have also been used for shaping in~\cite{Sommer-itw09,Ferdinand-twc16,di_pietro-com17,Zhou-commlett17,khodaiemehr-com17}, but never been used for shaping {C}onstruction D' lattices.

LDPC codes have been implemented in a wide variety of applications because of their capacity-achievability, efficient encoding, low-complexity decoding, and suitability for hardware implementation. For these reasons, LDPC codes are also suitable for constructing lattices. Recently Branco~da~Silva and Silva~\cite{da_silva-it19} proposed efficient encoding and decoding for {C}onstruction D’ lattices, particularly for LDPC codes. A codeword and cosets of component linear codes are used to form systematic codewords for {C}onstruction D' lattices. This encoding method naturally produces lattice points in a hypercube. 

However, hypercube does not provide shaping gain. A shaping lattice $\Ls$ is needed to do so. Let $\G$ and $\Gs$ be the generator matrix of $\Lc$ and $\Ls$, respectively. The check matrix of $\Lc$ is $\H=\G^{-1}$. To build a nested lattice code, $\Ls\subseteq\Lc$ must be satisfied~\cite[p.~179]{Zamir-2014}; this holds\textit{ iff }$\H\cdot\Gs\in\Zn$~\cite[Lemma 1]{kurkoski-it18}. To perform shaping, the mapping from an integer vector $\bb$ to a lattice point $\x'$ denoted $\x' = \G\bb-Q_{\Ls}(\G\bb)$~\cite[eq.~(21)]{kurkoski-it18} for a shaping lattice quantizer $Q_{\Ls}$, can be achieved as long as the integers $b_i \in \{0,1,\ldots,M_i-1\}$ are selected with $M_i$ related to the diagonal elements of $\H$ and $\Gs$. However, the encoding and decoding method in~\cite{da_silva-it19} cannot be applied to non-hypercubical shaping.

To tackle the above problem, the main contributions of this paper are as follows. We provide a definition of {C}onstruction D' using check-matrix perspective, which is equivalent to the congruences perspective~\cite{Conway-1999}. We propose two encoding methods and a decoding algorithm for {C}onstruction D' suitable for power-constrained channels.  \emph{Encoding method A} encodes integers with an approximate lower triangular (ALT) check matrix. \emph{Encoding method B} shows how binary information bits are mapped to a lattice point using the check matrices of the underlying nested linear codes of a {C}onstruction D' lattice. We present a multistage successive cancellation decoding algorithm employing binary decoders. The \emph{re-encoding} mapping an estimated binary codeword to a lattice point is required during decoding, and this is consistent with encoding method B; these methods are distinct from~\cite{da_silva-it19}.  

Motivated by~\cite{chen-istc18}, we also construct QC-LDPC codes to form {C}onstruction D' lattices (termed LDPC code lattices), because QC-LDPC codes are widely used in recent wireless communication standards. A design of QC-LDPC code $\C_0$ with a parity-check matrix $\H_0$ is presented, where the position of non-zero blocks is found by binary linear programming. A \emph{subcode condition} $\C_0\subseteq\C_1$ must be satisfied to form a 2-level {C}onstruction D' lattice, and this is not straightforward. In~\cite{da_silva-it19}, $\H_0$ was obtained from $\H_1$ by performing check splitting or PEG-based check splitting. In contrast to~\cite{da_silva-it19} we design $\H_0$ and construct $\H_1$ from $\H_0$. Simulation results of using well-known low-dimensional lattices for shaping a 2304-dimensional LDPC code lattice are given, and shaping gains of $E_8$, $BW_{16}$ and Leech lattices can be preserved.

\textit{Notation} A tilde indicates a vector or matrix which has only 0s and 1s --- $\tx$ and $\tH$ are binary while $\x$ and $\H$ are not necessarily so.  Operations over the real numbers $\mathbb R$ are denoted $+,\cdot$ while operations over the binary field $\F_2$ are denoted $\oplus, \odot$. The matrix transpose is denoted $\left(\cdot\right)^\textrm{t}$. Element-wise rounding to the nearest integer is denoted $\lfloor \cdot \rceil$. 
   
\section{{C}onstruction D'} \label{section:Dprimelattice}
An $n$-dimensional lattice $\Lambda$ is a discrete additive subgroup of $\Rn$. Let a generator matrix of $\Lambda$ be $\G$ with basis vectors in columns. For integers $\bb \in \Zn$, a vector $\x$ is a lattice point given by $\x=\G\cdot\bb$. This is equivalent to $\H \cdot \x = \bb$, where the check matrix is $\H = \G^{-1}$.

Consider nested linear codes $\C_0 \subseteq \C_1 \subseteq \cdots \subseteq \C_a = \F_2^n$ for level $a\geq1$. Let an $n$-by-$n$ matrix of row vectors $\h_j = [h_1,\ldots,h_n]$ be denoted $\tH = [\h_1,\h_2,\ldots,\h_n]^\textrm{t}$ for $j=1,\ldots,n$. The parity-check matrix of $\C_i$ is $\tH_i$ for $i=0,1,\ldots,a-1$. The dimension of $\C_i$ is $k_i = n-m_i$, where $m_i$ is the number of rows in $\tH_i$, from $\h_{k_i + 1}$ to $\h_n$.

{C}onstruction D' converts a set of parity-checks defining nested linear codes into congruences for a lattice. 
\begin{definition}[{C}onstruction D' (congruences)]~\cite[p.~235]{Conway-1999}\label{definition:constructionDprimecongruences} 
Let $\C_0 \subseteq \C_1 \subseteq \cdots \subseteq \C_a = \F_2^n$ be nested linear codes. Let the dimension of $\C_i$ be $k_i$. Let $\h_1,\h_2,\ldots,\h_n$ be a basis for $\F_2^n$ such that $\C_i$ is defined by $n-k_i$ parity-check vectors $\h_{k_i+1},\ldots,\h_n$. Then the {C}onstruction D' lattice is the set of all vectors $\x \in \Zn$ satisfying the congruences:
\begin{align}
\h_j\cdot\x\equiv0 \quad (\text{mod } 2^{i+1}), \label{eqn:congruences}
\end{align}
for all $i \in \{0,\ldots,a-1\}$ and $k_i+1 \leq j \leq n$.
\end{definition}

\begin{definition}[{C}onstruction D' (check matrix)] \label{definition:constructionDprime}
Let a unimodular matrix $\tH$ be the check matrix of nested linear codes $\C_0 \subseteq \C_1 \subseteq \cdots \subseteq \C_a = \F_2^n$. The dimension of $\C_i$ is $k_i$. Let $\D$ be a diagonal matrix with entries $d_{j,j} = 2^{-i}$ for $k_{i-1} < j \leq k_i$ for $i = 0,1,\ldots,a$ where $ k_{-1}=0$ and $k_a =n$. Then the {C}onstruction D' lattice is the set of all vectors $\x \in \Zn$ satisfying: $\H\cdot\x$ are integers, where $\H = \D\cdot\tH$ is the lattice check matrix.
\end{definition}

If $\tH$ is unimodular, then the {C}onstruction D' lattice $\Lambda \subset \Zn$. To see this, $\G = \H^{-1} = \tH^{-1}\cdot\D^{-1}$. Since $\tH$ is unimodular, $\tH^{-1}$ is integer. $\D^{-1}$ also is integer, thus $\G$ is an integer matrix. As a matter of design, after $\tH_0$ to $\tH_{a-1}$ are fixed, the upper rows of $\tH$ should be chosen such that $\tH$ is unimodular; it is also convenient to choose these upper rows so that $\tH$ is ALT form. The two definitions are equivalent if the congruence~(\ref{eqn:congruences}) is expressed: $\h'_j \cdot\x$ is an integer, for $\h'_j = \h_j /  2^{i+1}$ and $\x \in \Zn$.

The volume-to-noise ratio (VNR) is conventionally used while measuring error-correction performance of lattices, and can be defined $\VNR = V^{2/n}(\Lambda)/2\pi e \var$, where the volume of an $n$-dimensional {C}onstruction D' lattice is $V(\Lambda)=2^{an-\sum_{i=0}^{a-1}k_i}$, so that $\VNR$ is the distance to the Poltyrev limit.

\section{Encode and Decode {C}onstruction D'} \label{section:encodedecode}
Two equivalent encoding methods and a decoding algorithm for {C}onstruction D' lattices $\Lambda$ are given in this section. The encoding describes explicitly the mapping from bits to a lattice point. The connection between a lattice point and the modulo-value of lattice component (in short, a codeword of binary code $\C_i$) makes it possible to decode using an optimal decoder of $\C_i$. Encoding method A finds a lattice point of $\Lambda$ using its check matrix $\H$. Encoding method B illustrates how bits are mapped to integers, and accordingly lattice components are produced by check matrix $\tH$ of nested linear codes. The decoding algorithm is then given in Subsection~\ref{subsection:decode}.

\subsection{Encoding Method A} \label{subsection:encodeA}
Near linear-time encoding of LDPC codes can be accomplished using check matrix in the ALT form~\cite{richardson-it01*3}. This idea inspired us to implement encoding of {C}onstruction D' lattice $\Lambda$ with a similar procedure. The steps are distinct from~\cite{richardson-it01*3} because check matrix $\H$ of $\Lambda$ is a real-valued square matrix.

Integers $\bb$ are provided, and the corresponding lattice point $\x$ is found by solving: $\H\cdot\x=\bb$. If $\H$ is not too big, then $\x$ can be found by matrix inversion: $\x = \H^{-1}\cdot\bb$. If $\H$ is large but is sparse and in the ALT form, as may be expected for {C}onstruction D' lattices based on LDPC codes, then the following procedure can be used.

Suppose that $\H$ is in the ALT form, that is, it is partially lower triangular. Specifically, $\H$ can be written as:
\begin{align}
\H = 
\renewcommand\arraystretch{1}
\begin{bmatrix}
\mB &\T \\
\mX &\mC
\end{bmatrix}
\end{align}
where $\T$ is an $s$-by-$s$ lower-triangular matrix with non-zero elements on the diagonal; $\mX$ is a $g$-by-$g$ square matrix. The "gap" is $g$---the smaller the gap, the easier the encoding. Let $\Delta = (\mX-\mC\T^{-1}\mB)^{-1}$. The blockwise inverse of $\H$ is:
\begin{align}
\H^{-1} = 
\renewcommand\arraystretch{1.2}
\begin{bmatrix}
-\Delta\mC\T^{-1}            &\Delta       \\
\T^{-1}+\T^{-1}\mB\Delta\mC\T^{-1} &-\T^{-1}\mB\Delta
\end{bmatrix}
\label{eqn:inverseH}
\end{align}

Using the block structure, $\H\cdot\x=\bb$ can be written as:
\begin{align}
\renewcommand\arraystretch{1}
\begin{bmatrix}
\mB &\T \\
\mX &\mC
\end{bmatrix}
\cdot \renewcommand\arraystretch{1}
\begin{bmatrix}
x_1   \\
\vphantom{\int^0}\smash[t]{\vdots}\\
x_g   \\
x_{g+1} \\
\vphantom{\int^0}\smash[t]{\vdots}\\
x_n
\end{bmatrix}
= 
\renewcommand\arraystretch{1}
\begin{bmatrix}
b_1   \\
\vphantom{\int^0}\smash{\vdots}\\
b_g   \\
b_{g+1} \\
\vphantom{\int^0}\smash{\vdots}\\
b_n
\end{bmatrix}
\end{align}
To perform encoding, first $x_1,\ldots,x_g$ are found using~(\ref{eqn:inverseH}):
\begin{align}
\renewcommand\arraystretch{1}
\begin{bmatrix}
x_1   \\
\vphantom{\int^0}\smash{\vdots}\\
x_g   
\end{bmatrix}
=
\begin{bmatrix}
-\Delta\mC\T^{-1}  &\Delta
\end{bmatrix}
\cdot \bb.
\label{eqn:encodeAstep1}
\end{align}
Then, coordinates $x_{g+1},\ldots,x_n$ are found sequentially by back-substitution, using the lower triangular structure. For $i=g+1,\ldots,n$:
\begin{align}
x_i = \frac{1}{h_{j,i}}\bigg(b_j - \sum_{l=1}^{i-1} h_{j,l}x_l\bigg)
\label{eqn:encodeAstep2}
\end{align}
where $j=i-g$.

This method is efficient when $g$ is small and $\H$ is sparse. It uses pre-computation and storage of the $g$-by-$g$ matrix in~(\ref{eqn:encodeAstep1}). The sum in~(\ref{eqn:encodeAstep2}) is performed over non-zero terms, so few terms appear for sparse $\H$. If the parity-check matrix $\H$ is purely triangular, then encoding is simply performed by back-substitution.

\subsection{Encoding Method B}  \label{subsection:encodeB}
Encoding can also be performed using information vectors $\bu_i \in \F_2^{k_i}$ of $\C_i$ for $i=0,1,\ldots,a$ and $\bu_a=\z\in\Zn$. In this method, we show explicitly how $\bu_i$ and corresponding integers $\bb$ are related to a lattice point $\x$.

For clarity, consider $a = 3$. The integers $\bb$ are related to $\bu_0, \bu_1, \bu_2$ and $\z$ as:
\begin{equationarray}{crl}
b_j = &u_{0_j} +2u_{1_j} + 4u_{2_j} + 8z_j  &\text{ for $\ 1\leq j\leq k_0$}\\
b_j = &         u_{1_j}  + 2u_{2_j} + 4z_j  &\text{ for $k_0<j\leq k_1$} \\
b_j = &                     u_{2_j} + 2z_j  &\text{ for $k_1<j\leq k_2$} \\
b_j = &                                z_j  &\text{ for $k_2<j\leq n$}                           
\end{equationarray}    

Let $\bu_i'$ be the zero-padded version of $\bu_i$, to have $n$ components:
\begin{align}
\bu_i' = [u_{i_1},u_{i_2},\dots,u_{i_{k_i}},\underbrace{0,\dots,0}_{n-k_i}]^\textrm{t}
\end{align}
Then, the integer vector $\bb$ is written as:
\begin{align}
\bb = \D \cdot(\bu_0'+2\bu_1'+4\bu_2'+8\z), \label{eqn:b}
\end{align}
where $\D$ is given Definition~\ref{definition:constructionDprime}. 

For {C}onstruction D', the lattice point $\x$ may be decomposed as: 
\begin{align}
\x = \x_0+2\x_1+\cdots+2^{a-1}\x_{a-1}+2^a\x_a,\label{eqn:xdecomposition}
\end{align}
with components $\x_i$ depending on $\bu_i$ expressed below; $\x_i$ are not necessarily binary. 

Now we describe how information bits are related to a lattice point, and show that recovering integers from a lattice point is possible. Using $\H\cdot\x=\bb$, $\H = \D\cdot\tH$ and~(\ref{eqn:b})--(\ref{eqn:xdecomposition}) we have
\begin{align}
    \H \cdot \x                      &=\bb  \label{eqn:connectxandu}\\
    \tH \cdot \x                     &=\D^{-1} \cdot \bb \\
     \tH \cdot(\x_0+2\x_1+\cdots+2^a\x_a)  &= \bu_0'+2\bu_1'+\cdots+2^a\z
\end{align}
and the lattice components $\x_i \in \Zn$ satisfy:
\begin{align}
&\tH \cdot \x_i = \bu_i'  \quad \text{for } i=0,\dots,a-1   \quad \text{and} \label{eqn:encstep1} \\
&\tH \cdot \x_a = \z    \label{eqn:encstep2}
\end{align}
Note that encoding performed using~(\ref{eqn:encstep1})--(\ref{eqn:encstep2}) is equivalent to encoding method A in the previous subsection.

\subsection{Decoding {C}onstruction D'} \label{subsection:decode}

The multistage decoding for {C}onstruction D lattices first introduced in~\cite{vem-isit14} can produce an estimate of information bits $\hu_i$ from a binary decoder of $\C_i$, and the estimated component $\hx_i$ is obtained from $\hu_i$ by re-encoding $\hx_i = \G_i\cdot\hu_i$~\cite{ matsumine-glocom18}. An optimal decoder of $\C_i$ provides low complexity decoding. Recently multistage decoding similar to~\cite{vem-isit14} and~\cite{ matsumine-glocom18} was proposed for {C}onstruction D'~\cite{da_silva-it19}, including re-encoding steps to compute the cosets using estimate of lattice component of all previous levels. 

\begin{figure}[t]
\centering
\includegraphics[width=0.48\textwidth]{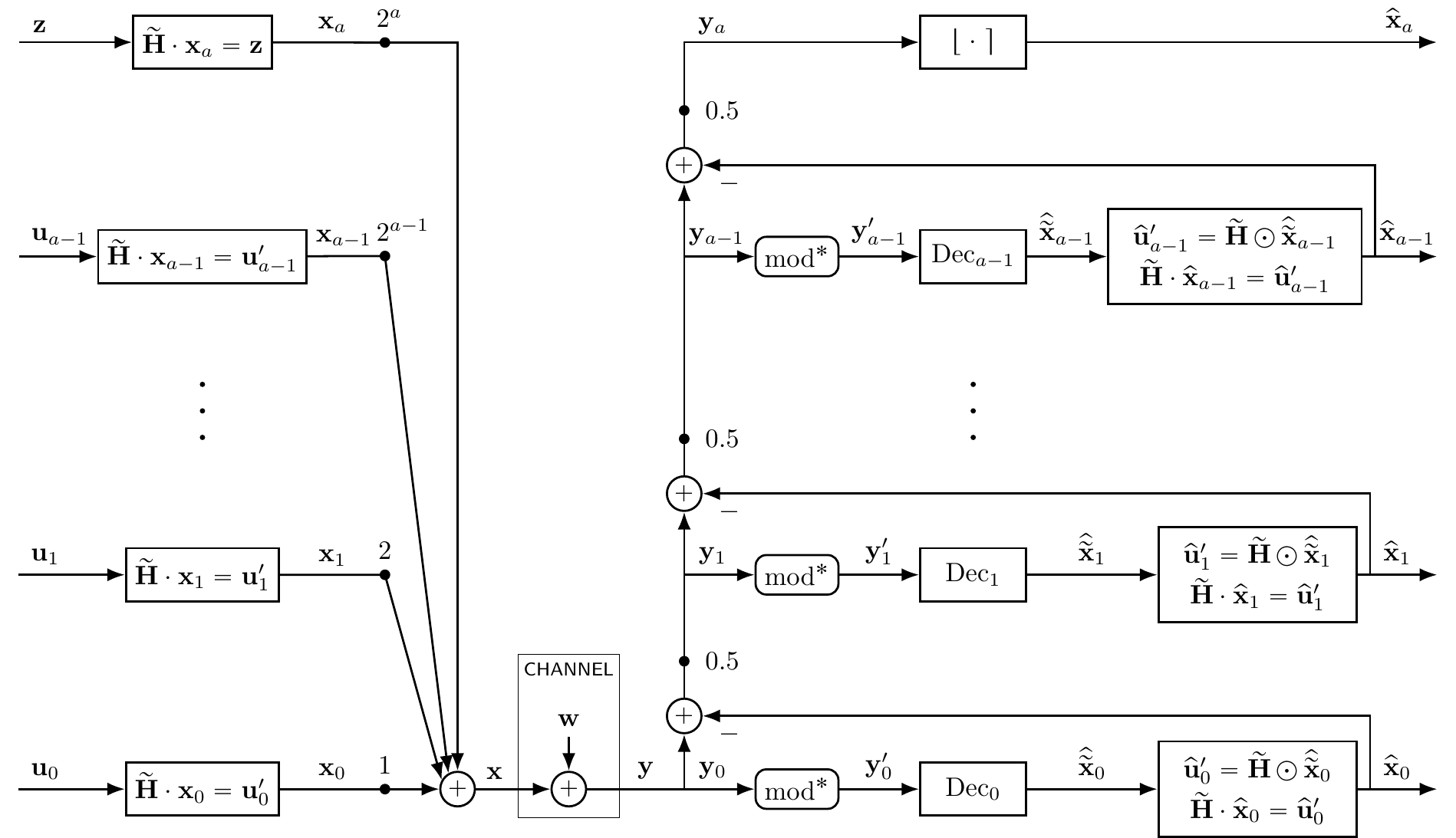}
\caption{Block diagram of proposed encoding and decoding {C}onstruction D' lattices. $\starmod$ denotes the "triangle-function" $\starmod(\y_i) = \left|\modtwo(\y_i+1)-1\right|$ where $\textrm{mod}_2$ indicates a modulo-2 operation.}
\label{fig:blockdiagramDprimelattice}
\end{figure}

We propose a multistage successive cancellation decoding algorithm suitable for {C}onstruction D' coding lattices to be used with shaping lattices, likewise employing a binary decoder $\Dec_i$ of $\C_i$. The encoding and decoding scheme of this paper is shown in Fig.~\ref{fig:blockdiagramDprimelattice}, where encoding method B is to demonstrate the validity of the decoding algorithm.  

\begin{proposition}
For {C}onstruction D', the lattice component $\x_i$ is congruent modulo 2 to a codeword $\tx_i\in\C_i$, for $i=0,\ldots,a-1$.
\end{proposition}
\begin{IEEEproof}
The lattice component $\x_i$ satisfies $\tH\cdot\x_i=\bu_i'$ and the codeword satisfies $\tH_i\odot\tx_i=\zero$. Recall the last $n-k_i$ positions of $\bu_i'$ are 0s. Row $l$ of $\tH_i$ is equal to row $l+k_i$ of $\tH$, call this row $\h_l$. By definition, $\h_l\cdot\x_i = 0$ and $\h_l\odot\tx_i=0$ for $l=1,2,\ldots,n-k_i$. Thus, $\x_i \mod 2=\tx_i$ and the proposition holds.
\end{IEEEproof}

Consider a lattice point $\x$ transmitted over a channel and the received sequence is $\y_0=\x+\w$, where $\w$ is noise. Decoding proceeds recursively for $i=0,1,\ldots,a-1$. The decoding result at level $i-1$ is used before beginning decoding at level $i$. Given $\y'_i \in [0,1]$, the decoder $\Dec_i$ produces a binary codeword $\thx_i$ closest to $\y'_i$, which is an estimate of $\tx_i$. It is necessary to find $\hx_i$. If $\tx_i$ is not systematic, first find $\hu_i' = \tH \odot \thx_i$. Then re-encoding is performed to find $\hx_i$, that is,~(\ref{eqn:encstep1}). This estimated component $\hx_i$ is subtracted from the input, and this is divided over reals by 2: $\y_{i+1} = (\y_i-\hx_i)/2$ to form $\y_{i+1}$, which is passed as input to the next level. This process continues recursively, until $\y_a$ is obtained. The integers are estimated as $\hx_a = \lfloor \y_a \rceil$. The estimated lattice point is written as $\hx = \hx_0+2\hx_1+\cdots+2^a\hx_a$.

\begin{table*}[t]
\caption{Prototype matrix of $\H_0$ with circulant size $Z=96$ and block length $n=2304$. $*$ denotes a double circulant.}
\label{table:H0}
\centering
\renewcommand\arraystretch{0.9}
\resizebox{\textwidth}{!}{
\begin{tabular}{*{22}{r}cr}                                                                                                                                                    
-1 & -1 & 53 & -1 & 15 & 56 & -1 & -1 & 55 & 35 & -1 & 8 & 0 & -1 & -1 & -1 & -1 & -1 & -1 & -1 & -1 & -1 & -1 & -1 \\ 
                                                                                                                
-1 & -1 & 26 & -1 & -1 & 51 & -1 & 59 & 14 & -1 & 16 & -1 & 0 & 0 & -1 & -1 & -1 & -1 & -1 & -1 & -1 & -1 & -1 & -1 \\ 
                                                                                                                
18 & -1 & 3 & -1 & -1 & 82 & 42 & -1 & 33 & -1 & -1 & -1 & -1 & 0 & 0 & -1 & -1 & -1 & -1 & -1 & -1 & -1 & -1 & -1 \\  
                                                                                                                
-1 & 30 & 73 & 53 & -1 & 49 & -1 & -1 & 8 & -1 & -1 & -1 & -1 & -1 & 0 & 0 & -1 & -1 & -1 & -1 & -1 & -1 & -1 & -1 \\  
                                                                                                                
-1 & 67 & -1 & 15 & 84 & -1 & -1 & -1 & -1 & -1 & -1 & -1 & -1 & 3 & -1 & 82 & 0 & -1 & -1 & -1 & -1 & -1 & -1 & -1 \\ 
                                                                                                               
-1 & -1 & -1 & -1 & -1 & 71 & 83 & 34 & -1 & -1 & -1 & -1 & -1 & 0 & -1 & -1 & 25 & 0 & -1 & -1 & -1 & -1 & -1 & -1 \\ 
                                                                                                               
-1 & -1 & -1 & -1 & -1 & -1 & 8 & 27 & 87 & -1 & -1 & -1 & 0 & -1 & -1 & -1 & -1 & 59 & 0 & -1 & -1 & -1 & -1 & -1 \\  
                                                                                                               
-1 & -1 & -1 & -1 & -1 & -1 & -1 & -1 & 91 & -1 & 62 & 52 & -1 & -1 & -1 & 0 & -1 & -1 & 6 & 0 & -1 & -1 & -1 & -1 \\  
                                                                                                               
-1 & -1 & -1 & -1 & -1 & -1 & -1 & -1 & -1 & 11 & 5 & 17 & -1 & -1 & 0 & -1 & -1 & -1 & -1 & 12 & 0 & -1 & -1 & -1 \\  
                                                                                                               
-1 & -1 & 2 & 43 & 53 & -1 & -1 & -1 & -1 & -1 & -1 & -1 & -1 & -1 & 73 & -1 & -1 & -1 & -1 & -1 & 34 & 0 & -1 & -1 \\ 
                                                                                                                
54 & -1 & 26 & -1 & -1 & 12 & -1 & -1 & -1 & -1 & -1 & -1 & -1 & -1 & -1 & -1 & -1 & -1 & -1 & -1 & -1 & 9 &\ 0 & 0 \\                                                                                                                
52 & 91 & -1 & -1 & -1 & -1 & -1 & -1 & -1 & 38 & -1 & -1 & 13 & -1 & -1 & -1 & -1 & -1 & -1 & -1 & -1 & -1 & $66/71^*$  & 0 \\
\end{tabular}%
}
\end{table*}
\begin{table*}[t]
\caption{Prototype matrix of $\H_1$ with circulant size $Z=96$ and block length $n=2304$. $*$ denotes a double circulant.}
\label{table:H1}
\centering
\renewcommand\arraystretch{0.9}
\resizebox{\textwidth}{!}{
\begin{tabular}{*{22}{r}cr}  
54 & 67 & 26 & 15 & 84 & 12 & 8 & 27 & 87 & 11 & 5 & 17 & 0 & 3 & 0 & 82 & 0 & 59 & 0 & 12 & 0 & 9 & 0 & 0 \\                                                                                                            
52 & 91 & 2 & 43 & 53 & 71 & 83 & 34 & 91 & 38 & 62 & 52 & 13 & 0 & 73 & 0 & 25 & 0 &6 & 0 & 34 & 0 & $66/71^*$ & 0 \\  
\end{tabular}%
}
\end{table*}                           

\section{Design LDPC Code Lattices} \label{section:ldpcdesign}

Branco~da~Silva and Silva also addressed the design of multilevel LDPC code lattices~\cite{da_silva-it19}. Using row operations on the parity-check matrix $\mathbf{H}_0$ of the first level component code $\C_0$, a parity-check matrix that has the desired row and column degree distributions is obtained---the two check matrices both describe $\C_0$. In our work, QC-LDPC codes are designed using binary linear programming to guarantee that the necessary supercode can be constructed, as well as to satisfy the column and the row weight distribution. 

\subsection{{C}onstruction D' Lattices Formed by QC-LDPC Codes}

The parity-check matrix $\mathbf{H}_0$ is expressed as
\begin{eqnarray}
\mathbf{H}_0&=&\begin{aligned}
\renewcommand\arraystretch{1}
\begin{bmatrix} 
\mathbf P^{p^{}_{1,1}} &  \mathbf P^{p^{}_{1,2}} & \cdots & \mathbf P^{p^{}_{1,N}} \\ 
\mathbf P^{p^{}_{2,1}} & \mathbf P^{p^{}_{2,2}} & \cdots & \mathbf P^{p^{}_{2,N}} \\
\vdots & \vdots & \ddots & \vdots \\ 
\mathbf P^{p^{}_{M,1}} & \mathbf P^{p^{}_{M,2}} & \cdots & \mathbf P^{p^{}_{M,N}}
\end{bmatrix},   \label{eqn:prototypematrix}
\end{aligned}
\end{eqnarray}
where $\mathbf P^{p^{}_{i,j}}$ is a $Z$-by-$Z$ matrix (or a circulant) corresponding to the element $p^{}_{i,j}$ of a prototype parity-check matrix. For $p^{}_{i,j}=-1$, instead use the all-zeros matrix. For $p^{}_{i,j}=0$, $\mathbf P$ is an identity matrix. And $\mathbf P$ is a right-shift cyclic-permutation matrix for an integer $0<p^{}_{i,j}<Z$ indicating the shift amount.

This paper proposes two-level {C}onstruction D' lattices based on QC-LDPC codes.  The first level component code $\mathcal{C}_0$ has an $M$-by-$N$ prototype parity-check matrix while the second level component code $\mathcal{C}_1$ is a $2$-by-$N$ prototype parity-check matrix. The code $\mathcal{C}_0$ has a design rate $1-M/N$, and $\mathcal{C}_1$ is a high-rate code---a column weight 2, row weight $N$ parity-check matrix is sufficient; column weight 2 was also used in~\cite{chen-istc18}. The code $\mathcal{C}_0$ is a subcode of $\mathcal{C}_1$ thus the prototype matrix of $\mathcal{C}_1$ is a matrix obtained from linear combinations of a $\mathcal{C}_0$ prototype submatrix. Binary linear codes $\C_0$ and $\C_1$, and their parity-check matrices $\H_0$ and $\H_1$ are nested.

The parity-check matrix $\mathbf{H}_1$ does not provide good row and column distributions if the rows were taken from $\mathbf{H}_0$, thus we find $\H_1$ using linear combinations of rows in $\H_0$. Let the set $\mathcal A_q$ be block rows of $\mathbf H_0$ such that their sum is a single block row of weight $N$ and column weight 1, for $q=1,2$. In addition, $\mathcal A_1$ and $\mathcal A_2$ are disjoint.

The parity-check matrix $\mathbf{H}_1$ can be expressed as
\begin{eqnarray}
\mathbf{H}_1&=&\begin{bmatrix}\mathbf{H}_1'\\ \mathbf{H}_2' \end{bmatrix}, \label{eqn:obtainH1}
\end{eqnarray}
where $\mathbf{H}_1'$ and $\mathbf{H}_2'$ are the sum of block rows $\mathcal A_1$ and $\mathcal A_2$, respectively: 
\begin{eqnarray}
\mathbf{H}_q'&=&\bigoplus_{k\in \mathcal{A}_q}[\mathbf P^{p^{}_{k,1}} \; \mathbf P^{p^{}_{k,2}} \; \cdots \; \mathbf P^{p^{}_{k,N}}],
\end{eqnarray}
for $q=1,2$.  Accordingly, $\mathbf H_1$ is a QC-LDPC matrix with column weight 2.

\subsection{Binary Programming for Prototype Matrix Construction}

To form a two-level {Construction D'} lattice using QC-LDPC codes, the two component binary codes are needed to satisfy the properties given in the previous subsection. One part of the design is to find the location of the non-zero circulants.

The goal is to design a matrix, given several constraints: the subcode condition, row and column weight degree distributions, and the matrix should be in the ALT form to enable efficient encoding.  Binary linear programming can be used to satisfy these constraints to provide a desired prototype matrix~\cite{sulek-access19}.

Set up the programming problem by writing the $M \times N$ matrix as
\begin{align} \mathbf{A}=
\renewcommand\arraystretch{0.8}
\begin{bmatrix}
a^{}_{1,1} & a^{}_{1,2} & \cdots & a^{}_{1,N} \\
\vdots  & \vdots  & \ddots & \vdots   \\
a^{}_{M,1} & a^{}_{M,2} & \cdots & a^{}_{M,N} 
\end{bmatrix},  \label{eqn:binmatrix}
\end{align}
where $a^{}_{i,j} \in \{0,1\}$ is a binary variable and $a^{}_{i,j} = 1$ indicates a non-zero block.  The row weights are $\mathbf{r}=\{r_1, r_2, \cdots r_M\}$ and the column weights are $\mathbf{c}=\{c_1, c_2, \cdots c_N \}$.  There are $M$ row constraints and $N$ column constraints:
\begin{eqnarray}
\begin{matrix}
\textrm{Row $i$ has weight $r_i$}  &\Leftrightarrow  & a^{}_{i,1} + \cdots + a^{}_{i,N} & = &r_i  \\
\textrm{Col $j$ has weight $c_j$}  &\Leftrightarrow  & a^{}_{1,j} + \cdots + a^{}_{M,j} & = &c_j
\end{matrix} \label{eqn:constraint1}
\end{eqnarray}
for $i \in \{1,\ldots,M\}$ and $j \in \{1,\ldots,N\}$. For one of the subcode constraints, we want rows from $\mathcal A_q$ to sum to a single block row with weight $N$, so we add a constraint for $q=1,2$:
\begin{align}
    \sum_{i \in \mathcal{A}_q} \sum_{j=1}^{N} a^{}_{i,j}= N \label{eqn:constraint2}
\end{align}

Also, we want a constraint for the ALT form to force 1's along the offset-by-one diagonal with the constraint:
\begin{align}
    \sum_{i=1}^{M-1} a^{}_{i,N-M+1+i}=M-1, \label{eqn:constraint3}
\end{align}
in addition to another constraint to force all-zeros above the offset diagonal. 

The goal is to find $a^{}_{i,j}$ that satisfies the above conditions. This goal can be expressed using the following binary linear program:
\begin{align}
\min \sum_i \sum_j a^{}_{i,j}
\end{align}
subject to
\begin{eqnarray}
\label{eqn:matrixFormConstraints}
\mathbf{K} \cdot \mathbf{a}=
\begin{bmatrix}
\mathbf{r} & \mathbf{c} & N & N & M-1 & 0
\end{bmatrix}^\textrm{t},
\end{eqnarray}
where $\mathbf{K}$ is a constraint matrix that includes all the constraints described in (\ref{eqn:constraint1})--(\ref{eqn:constraint3}) and $\mathbf a$ is the vectorized version of $\mathbf A$. Because this is a binary programming problem, for the set $\mathcal A_q$, only one position will contain a 1 and the remaining $|\mathcal A_q| - 1$ positions will contain 0, in any column. The implementation of this optimization problem is easily solved using standard optimization packages. 

\subsection{Resulting Design}
\begin{figure}[t]
\centering
\includegraphics[width=0.428\textwidth]{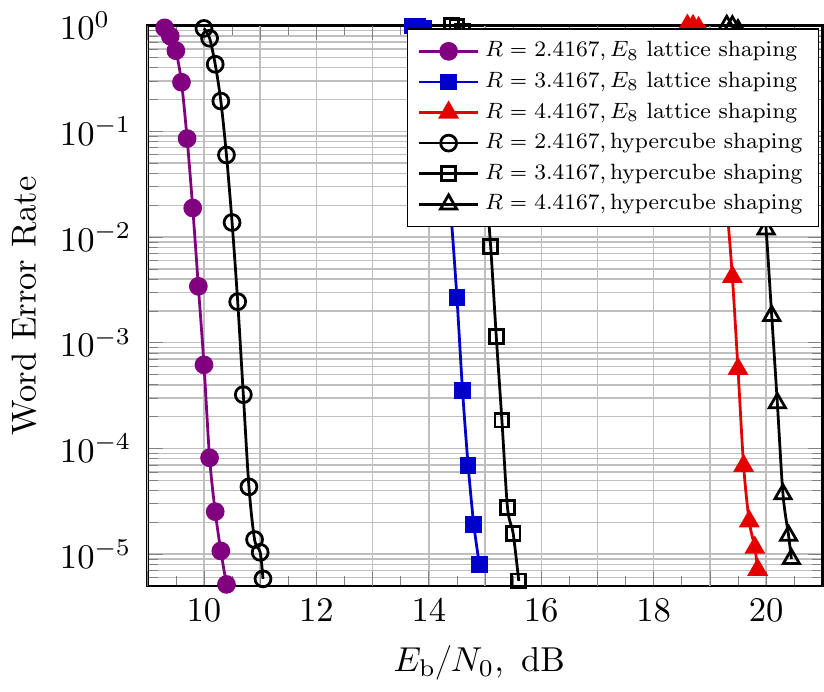}
\caption{Word error rate of shaping an $n=2304$-dimensional {C}onstruction D' lattice (formed by QC-LDPC codes) using $E_8$ lattice shaping and hypercube shaping.}
\label{fig:shapingresultlowrate}
\end{figure}
Now we give a specific design of binary QC-LDPC codes $\C_0$ and $\C_1$ for 2-level {C}onstruction D' lattices. The check matrices $\H_0$ and $\H_1$ are designed such that: 1) $\C_0 \subseteq \C_1$ 2) $\H_0$ and $\H_1$ are of full rank 3) $\H_0$ and $\H_1$ can be easily triangularized 4) $\H_0$ and $\H_1$ have girth as high as possible.

To meet the design requirements, first we use the binary linear programming in previous subsection to find a binary matrix $\mA$~(\ref{eqn:binmatrix}) with $M = 12$ rows and $N = 24$ columns, using degree distribution polynomials of variable nodes and check nodes: $\lambda(x) = \frac{1}{3}x+\frac{5}{12}x^2+\frac{1}{8}x^3+\frac{1}{8}x^5$ and $\rho(x)=\frac{2}{3}x^5+\frac{1}{3}x^6$, respectively, where $\lambda_dx^{d-1}$ and $\rho_dx^{d-1}$ means $\lambda_d$ and $\rho_d$ are the fraction of nodes with degree $d$. This structure is a modified version of~\cite[Table~I]{rosnes-it12}. The check matrix $\H_1$ is constructed~(\ref{eqn:obtainH1}) with $\mathcal A_1=\{5,7,9,11\}$ and $\mathcal A_2=\{6,8,10,12\}$. Using a circulant size $Z=96$, the prototype matrix as shown in Table~\ref{table:H0} can be generated by assigning $p^{}_{i,j}=-1$ for $a^{}_{i,j}=0$ and choosing the powers $-1<p^{}_{i,j}<Z$ for $a^{}_{i,j}=1$ such that $\H_0$ and $\H_1$ are free of 4-cycles and 6-cycles, where $\H_0$ can be constructed using~(\ref{eqn:prototypematrix}). The resulting $\H_1$ lifted from  Table~\ref{table:H1} has degree distribution polynomials $\lambda(X) = X$ and $\rho(X)=X^{23}$. The designed QC-LDPC codes $\C_0$ and $\C_1$ are of block length $n=2304$, with code rate $k_0/n=1/2$ and $k_1/n=11/12$, respectively.

The check matrix $\H$ of an LDPC code lattice can be then constructed from $\H_0$ and $\H_1$ using Definition~\ref{definition:constructionDprime}. Note that we did not use offset diagonal and we assigned a double circulant $p^*_{12,23}$ such that $\H_0$ and $\H_1$ can be easily triangularized. This provides efficient encoding and indexing~\cite{kurkoski-it18}.

\section{Numerical Results}

\begin{figure}[t]
\centering
\includegraphics[width=0.45\textwidth]{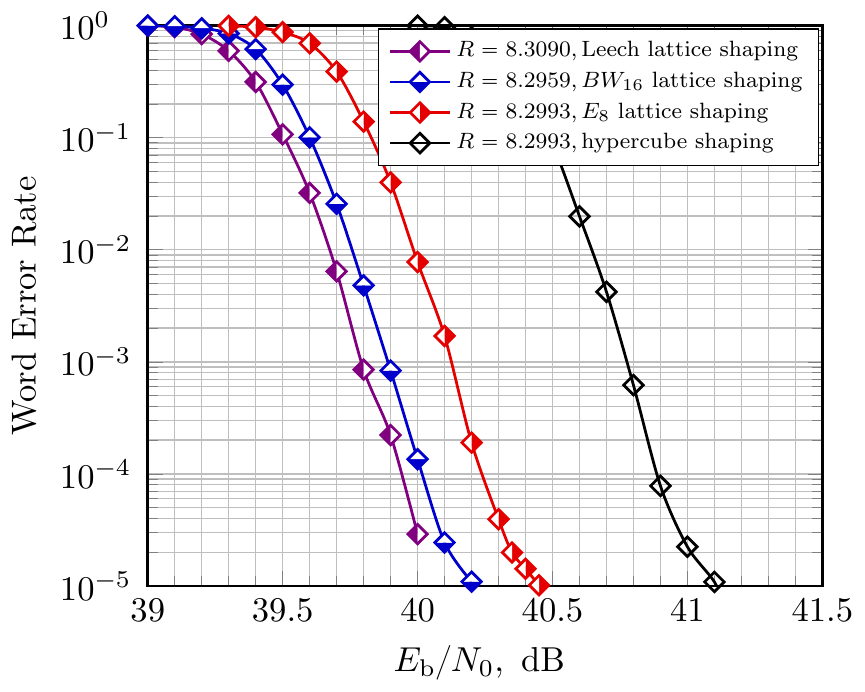}
\caption{Word error rate of shaping the 2304-dimensional {C}onstruction D' lattice using $E_8$, $BW_{16}$ and Leech lattice shaping. Hypercube shaping was performed at a close code rate for comparison.}
\label{fig:shapingresulthighrate}
\end{figure}

The $2304$-dimensional LDPC code lattice was evaluated on the power-constrained AWGN channel. The transmitted power can be reduced by {V}oronoi constellations using low-dimensional lattices, $E_8$, $BW_{16}$ and Leech lattices are considered in this paper. The direct sum of scaled copies (by a factor $K$) of a low-dimensional lattice produces a $2304$-dimensional shaping lattice $\Ls$ adapted to the proposed LDPC code lattice $\Lc$---$K$ is chosen such that the two lattices satisfy $\Ls\subseteq\Lc$, and thus form a nested lattice code to be used with dithering and MMSE scaling~\cite{Erez-it04}. For more details about shaping high-dimensional $\Lc$ using low-dimensional lattices, see~\cite{Ferdinand-twc16},~\cite{di_pietro-com17}. Rectangular encoding and its inverse indexing can be efficiently implemented~\cite{kurkoski-it18}. By choosing various $K$ we generated a variety of nested lattice codes with code rate $R$. For comparison, hypercube shaping was performed where lattice points were transformed into a hypercube $\B = \{0,1,\ldots,L-1\}^n$ for an even integer $L$. The belief propagation decoder of LDPC codes ran maximum 50 iterations.

The same code rate for both $E_8$ lattice shaping and hypercube shaping can be easily achieved. The word error rate using $K_{E_8} = L = 8,16,32$ is shown in Fig.~\ref{fig:shapingresultlowrate} as a function of $\ebnoshort = \SNR/2R$ with conventional signal-to-noise ratio (SNR), suggesting a shaping gain of $0.65\dB$. Let $K_{BW_{16}} = 280\sqrt{2}$ and $K_{\textrm{Leech}} = 168\sqrt{8}$, then $BW_{16}$ and Leech lattice shaping produce code rate approximately 8.2959 and 8.3090, respectively, close to $R=8.2993$ of choosing $K_{E_8} = L = 472$. Numerical results are given in Fig.~\ref{fig:shapingresulthighrate}. If we take account of the code rate differences, a $0.65\dB$, $0.86\dB$ and $1.03\dB$ shaping gain is preserved respectively, as the full shaping gain of $E_8$, $BW_{16}$ and Leech lattices.


\bibliographystyle{bibtex/IEEEtrannourl} 

\end{document}